\documentstyle[12pt,fleqn]{article}
\input epsf
\begin{document}
\newcommand{\mpl}{M_{\mbox{Pl}}}

\begin{titlepage}
\null\vspace{-72pt}
\begin{flushright}
{\footnotesize
FERMILAB-Pub-98/295--A\\
hep-ph/9809489\\
September 1998 \\
}
\end{flushright}
\renewcommand{\thefootnote}{\fnsymbol{footnote}}
\vspace{0.15in}
\baselineskip=24pt

\begin{center}
{\Large \bf Classical inflaton field induced creation of superheavy
dark matter}
\baselineskip=14pt
\vspace{0.75cm}

Daniel J.\ H.\ Chung\footnote{Electronic mail:
                {\tt djchung@feynman.physics.lsa.umich.edu}}\\
{\em Department of Physics and Enrico Fermi Institute\\
The University of Chicago, Chicago, Ilinois~~60637, and\\
NASA/Fermilab Astrophysics Center\\
Fermi National Accelerator Laboratory, Batavia, Illinois~~60510}\\
\end{center}
\baselineskip=24pt

\begin{quote}
\hspace*{2em}
We calculate analytically and numerically the production of superheavy
dark matter ($X$) when it is coupled to the inflaton field $\phi$
within the context of a slow-roll $m_\phi^2 \phi^2/2$ inflationary
model with coupling $g^2 X^2 \phi^2/2$.  We find that $X$
particles with a mass as large as 1000 $H_i$, where $H_i$ is the value
of the Hubble expansion rate at the end of inflation, can be produced
in sufficient abundance to be cosmologically significant today.  This
means that superheavy dark matter may have a mass of up to $10^{-3}
\mpl$.  We also derive a simple formula that can be used to estimate
particle
production as a result of a quantum field's interaction with a general
class of homogeneous classical fields. 
Finally, we note that the combined effect of the inflaton field and
the gravitational field on the $X$ field causes the production to be a
nonmonotonic function of $g^2$.
\vspace*{8pt}

PACS number(s): 98.80.Cq, 4.62.+v, 95.35.+d

\renewcommand{\thefootnote}{\arabic{footnote}}
\addtocounter{footnote}{-1}
\end{quote}
\end{titlepage}

\newpage

\baselineskip=24pt
\renewcommand{\baselinestretch}{1.5}
\footnotesep=14pt

\def\la{\mathrel{\mathpalette\fun <}}
\def\ga{\mathrel{\mathpalette\fun >}}
\newcommand{\tend}{t_e}
\newcommand{\omhsq}{\Omega_X h^2/S}
\newcommand{\gparam}{g M_{\mbox{Pl}}/H_i}
\newcommand{\omk}{\omega_k}
\newcommand{\ett}{\tilde{\eta}}
\newcommand{\aeff}{a_{\mbox{eff}}(r)}
\newcommand{\heff}{\protect{H_{\mbox{eff}}(r)}}
\newcommand{\heffsq}{\protect{H^2_{\mbox{eff}}(r)}}
\newcommand{\reff}{\protect{R_{\mbox{eff}}(r)}}

\def\question#1{{{\marginpar{\small \sc #1}}}}
\def\mpl{{{M_{\mbox{Pl}}}}}
\def\oph{{{\Omega_{\widetilde{\gamma}}h^2}}}
\def\be{\begin{equation}}
\def\ee{\end{equation}}
\def\ba{\begin{eqnarray}}
\def\ea{\end{eqnarray}}
\def\la{\mathrel{\mathpalette\fun <}}
\def\ga{\mathrel{\mathpalette\fun >}}
\def\fun#1#2{\lower3.6pt\vbox{\baselineskip0pt\lineskip.9pt
        \ialign{$\mathsurround=0pt#1\hfill##\hfil$\crcr#2\crcr\sim\crcr}}}
\def\pho{{{\widetilde{\gamma}}}}
\def\r0{{{R^0}}}
\def\rpi{{{R_\pi}}}
\def\glu{{{\widetilde{g}}}}
\def\sec{{{\mbox{sec}}}}
\def\GeV{{{\mbox{GeV}}}}
\def\MeV{{{\mbox{MeV}}}}
\def\SUSY{{{{\sc susy}}}}
\def\LSP{{{{\sc lsp}}}}
\def\LEP{{{{\sc lep}}}}
\def\LROCS{{{{\sc lrocs}}}}
\def\WIMPS{{{{\sc wimps}}}}
\def\cm{{{\mbox{cm}}}}
\def\photino{{{\mbox{photino}}}}
\def\gluino{{{\mbox{gluino}}}}
\def\mb{{{\mbox{mb}}}}
\def\avg#1{{{{\langle #1 \rangle }}}}
\def\taun{{{\tau_{9}}}}
\def\re#1{{[\ref{#1}]}}
\def\eqr#1{{Eq.\ (\ref{#1})}}
\def\mst{{{M_{\widetilde{S}}}}}
\newcommand{\fignum}{\refstepcounter{figure}\thefigure}
\newcommand{\intcvt}{$\pho-\r0$ conversion }
\newcommand{\sigv}{\avg{\sigma_{\r0 \pi^\pm \rightarrow \pho \pi^\pm} v}}

\vspace{36pt}
\centerline{\bf I. INTRODUCTION}
\vspace{24pt}

The rotation curves deduced from observing luminous matter (see for
example \cite{begeman}) indicate dark matter (DM) exists around
galaxies.  Furthermore, comparison of the peculiar velocities of many
galaxies with the detailed maps of density contrast suggest \cite{dekel} that
$\Omega > 0.3$.  However, these cannot be all in the form of baryons
according to the standard scenarios of big bang nucleosynthesis
\cite{olive,cst}.  Structure formation studies indicate that relativistic
dark matter is unlikely to make up most of the DM \cite{white}.  These
evidences suggest the existence of a cosmologically significant
abundance of nonbaryonic weakly interacting massive particles (WIMPs).
Since SUSY models (including string inspired ones) generically predict
new stable weakly interacting particles, the existence of WIMPs is
even more likely.

Despite the fact that the nature of the DM is still unknown, it is
usually thought that DM particles cannot be too heavy.  If the WIMP is
a thermal relic, then it was once in local thermodynamic equilibrium
(LTE) in the early universe, and its present abundance is determined
by its self-annihilation cross section.  As argued by Griest and
Kamionkowski \cite{griestkam}, the self annihilation cross section has
an upper bound of $\sim 1/M_X^2$ from considerations of unitarity,
while the temperature at which freeze out occurs increases as the
cross section is decreased.  Hence, the assumption of LTE gives an
upper bound of about 500 TeV to the mass of the dark matter.  The
present abundance of non-thermal relics (those that never attained
LTE) is not determined by their self-annihilation cross section
because their final abundance is not simply determined by the usual
freeze out scenario.  An example of a non-thermal relic is
the axion, and the present axion abundance is determined by the
dynamics of the phase transition associated with symmetry breaking.
Non-thermal relics are typically very light, e.g., the axion mass is
expected to be in the range $ 10^{-5}$ to $10^{-2}$eV \cite{raf}.

However, nonthermal relics can have masses many orders of magnitude
larger than the electroweak scale and can evade the unitarity bound of
Ref.\ \cite{griestkam}.  These nonthermal DM particles have been
called superheavy dark matter (SDM) in Ref.\ \cite{ckr1}.  SDM
scenarios have been discussed in conjunction with various production
mechanisms (see for example \cite{ckr2} and references therein), with
the gravitational production mechanism being arguably the least fine
tuned \cite{ckr1,kuzmin,damour}.  In this paper, we will explore the
idea of \cite{earlyidea}, which is to produce SDM by  the same mechanism
that is at work in what has been called ``preheating'' scenarios.

The main ingredient of the preheating scenarios, introduced in the
early 1990s, is the nonperturbative resonant transfer of energy to
particles induced by the coherently oscillating inflaton fields.  It
was realized that this nonperturbative mechanism can be much more
efficient than the usual perturbative mechanism for certain parameter
ranges of the theory \cite{traschen, koflindstar1, koflindstar2,
preheating}.  The basic picture can be seen as follows.  Suppose we
have a scalar field $X$ with a coupling $ g^2 \phi^2 X^2$  where
$\phi$ is a homogeneous classical inflaton field.  The mode  
equation for $X$ field then can be written in terms of a redefined
variable $\chi_k \equiv X_k a^{3/2}$ as 
\be 
\ddot{\chi_k}(t) + (A + 2 q \cos(2t)) \chi_k(t)=0
\label{eq:mathieu}
\ee
where $A$ depends on the energy of the particle and $q$ depends on the
inflaton field oscillation amplitude.
When $A$ and $q$ are constants, this equation is usually referred to as
the Mathieu equation which exhibits resonant mode instability for certain
values of $A$ and $q$.  In an expanding universe, $A$ and $q$ will
vary in time, but if they vary slowly compared to the
frequency of oscillations, the effects of resonance will remain.   If
the mode occupation number for the $X$ 
particles is large, the number density per mode of the $X$ particles
will be proportional to 
$|\chi_k|^2$.  If $A$ and $q$ have the appropriate values for
resonance, $\chi_k$ will grow exponentially in time, and hence the 
number density will attain an exponential enhancement above the usual
perturbative decay.  This period of enhanced rate of energy transfer
has been called preheating 
primarily because the particles that are produced during this period
have yet to achieve thermal equilibrium.

This resonant amplification leads to an efficient transfer of energy
from inflatons to other particles which may have stronger coupling to
other particles than the inflaton, thereby speeding up the reheating
process and leading to a higher reheating temperature than in the
usual scenario.  Another interesting feature is that particles of mass
larger than the inflaton mass can be produced through this coherent
resonant effect.  Such a process is negligible in a conventional
scenario of reheating \cite{usual}.  This has been exploited to
construct a baryogenesis scenario \cite{earlyidea} in which the baryon
number violating bosons with masses larger than the inflaton mass are
created through the resonance mechanism.  A natural
variation on this idea is to produce SDM by the same resonance
mechanism \cite{earlyidea}.    

Interestingly enough, what we find in our work is that in the context
of a slow-roll inflation with the potential $V(\phi)=m_\phi^2
\phi^2/2$ with the inflaton coupling of $g^2 \phi^2 X^2/2$, the
resonance phenomenon is mostly irrelevant to the production of SDM
because too many particles are produced when the resonance is
effective.  For the tiny amount of energy conversion needed for SDM
production (tiny means $\sim 10^{-17}$ of the total energy), the
coupling $g^2$ must be small enough (for a fixed $M_X$) such that
the motion of the inflaton field only at the transition out of the
inflationary phase generates just enough nonadiabaticity in the
mode frequency to produce SDM.  The rest of the oscillations, damped
by the expansion of the universe, will not contribute significantly to
particle production as in the resonant case.  In other words, the
quasi-periodicity necessary for a true resonance phenomenon is hardly
existent for the case when only an extremely tiny fraction of the
energy density is converted into SDM.  Of course, if the energy scales
are lowered such that a fair fraction of the energy density can be
converted to DM without overclosing the universe, this argument may
not apply.  However, in this paper, 
we will be mostly interested in producing SDM with masses larger than
the inflaton mass within the context of a large-field inflationary
scenario, where this argument will apply.  For the study of cases in
which the resonance starts to become efficient, we refer the reader to
to Refs.\ \cite{ivaylo,koflindstar2,traschen,earlyidea} and references
therein.

The main findings of this work are the following:  We find that
superheavy dark matter with a mass as large as $10^3 H_i$, where $H_i$
is the value of the Hubble expansion rate at the end of inflation, can
be produced in sufficient abundance to be cosmologically significant
today.  Typically, $H_i$ can be as large as $10^{13}$ GeV, which means
that the dark matter may have masses of the order of the GUT scale.
In the process of finding this estimate, we derive a simple formula
(in the spirit of 
Ref.\ \cite{berezin}), \eqr{eq:goodapprox}, that can be used to
estimate particle production resulting from a general class of
interactions with a time varying homogeneous classical field
(including the gravitational field).  Finally we observe that coupling
$X$ to the inflaton field can actually decrease the amount of SDM
produced as a consequence of the inflaton field variation canceling
some of the nonadiabaticity of the expansion rate responsible for
gravitational production of SDM.

This paper is organized as follows.  In the next section we will
specify the model and the inflationary scenario in which our
estimations are carried out.  In section III, we derive a general
formula to estimate particle production when a
quantum field interacts weakly with a general class of homogeneous
classical fields.  We then compare our approximations to two exact
solutions.  Section IV follows where we 
apply the estimation to our model described in Section II.  We then
present numerical results for comparison and a
better estimation of the maximum cosmologically interesting SDM mass
in our model.  Finally, we conclude with a summary in Section V.


\vspace{36pt}
\centerline{\bf II. MODEL}
\vspace{24pt}

Two conditions are necessary for the viability of the SDM
scenario \cite{ckr1}: {\it a)}
their interaction rate must be sufficiently weak such that local
thermodynamic equilibrium (LTE) was never obtained
and {\it b)} the $X$ particles must be cosmologically stable.  As we
will see, 
because LTE necessitates the reaction rate to be larger than the
Hubble expansion rate while the reaction rate involves at least an
inverse mass squared suppression coming from the cross section
involved, large mass particles can naturally evade LTE.

Let us denote $\rho_X$ as the energy density of the SDM particles and
$n_X(t_e)$ as the number density of the SDM at time $t_e$ when
inflation ends.  As
shown in \cite{ckr1}, today's SDM density $\Omega_X\equiv
\rho_X(t_0)/\rho_C(t_0)$ (where $\rho_C(t_0)=3 H_0^2\mpl^2/8\pi$ and
$H_0=100\: h$ km sec$^{-1}$ Mpc$^{-1}$) can be expressed as
\be
\Omega_X h^2 \approx \Omega_R h^2\:
\left(\frac{T_{\mbox{RH}}}{T_0}\right)\:
\frac{8 \pi}{3} \left(\frac{M_X}{\mpl}\right)\:
\frac{n_X(t_{e})}{\mpl H_i^2}
\label{eq:omegachi}
\ee
where $H_i$ is the Hubble expansion rate at the end of inflation,
$T_0$ is the temperature today, 
$T_{\mbox{RH}}$ is the reheating temperature, and $\Omega_R h^2 \approx 4.31
\times 10^{-5}$ is the fraction of critical energy density that is in
radiation today.  Throughout this paper, we will give our results in
terms of $\omhsq$ where we have defined
\be
S \equiv (T_{\mbox{RH}}/(10^9 \mbox{GeV}))
(H_i/(10^{-6} \mpl))^2.
\ee
For a typical reheating
temperature of $10^9$ GeV, \eqr{eq:omegachi} implies that the SDM
energy density today will be $\Omega_X
h^2 \sim 10^{17} (\rho_X(t_e)/\rho(t_e))$ where $\rho(t_e)$ is the
total energy density at the end of inflation.  It
is indeed a very small fraction of the total energy density that needs to
be extracted to saturate the upper bound on the cosmological mass density.
Hence, the difficulty of our scenario lies in creating very few
particles, if these are to be the SDM.  

Now, consider the nonthermalization condition
\be
n_{X} \langle \sigma_A |v| \rangle \la H
\label{eq:equilib2}
\ee
which allows the 
evasion of  the unitarity upper bound on the mass of DM.
Using \eqr{eq:omegachi} with $\Omega_X h^2 < 1 $ and the fact that for
WIMPS, the averaged annihilation cross section  $\langle \sigma_A |v|
\rangle$ is less than $M_X^{-2}$ (unitarity bound), we can obtain the estimate 
\be
 \frac{n_X \langle \sigma_A |v| \rangle}{H_i} < \frac{ 7 \times
10^{-19}}{ (T_{\mbox{RH}}/10^9 \mbox{GeV})} \frac{(H_i/\mpl)}{(M_X/\mpl)^3}
\ee
which is the quantity that must be less than one at the end of
inflation to avoid thermalization.  If $H_i \approx 10^{-6} \mpl$
and 
\be
\left(\frac{M_X}{H_i}\right)\left(\frac{T_{\mbox{RH}}}{10^9
\mbox{GeV}}\right)^{1/3} > 10^{-2}, 
\ee
there is no LTE, and the particles density will evolve trivially as was
assumed in \eqr{eq:omegachi}.  Thus, because of the $M_X^{-2}$
generically coming from the cross section, SDM will generally
fail to achieve LTE irrespective of the exact value of the weak
coupling constant.   Note that this is a rather
conservative estimate since the reheating temperature is likely to be
larger and the cross section is likely to be smaller.  We also remark
that because the reheating temperature is usually much smaller
than the $X$ mass in SDM scenarios, the thermal production of the $X$
particles is usually negligible.\footnote{Since for times larger than
$t_e$, the interaction rate continues to be smaller than $H$, the
particles will not thermalize later either.}

For the $X$ particles to serve as DM, they must have a lifetime that is
longer than the age of the universe and be extremely massive.  One
possible source of SDM is the secluded and the messenger sectors of
the gauge mediated SUSY breaking models where SUSY can be broken at a
large scale (giving rise to large masses) while the secluded and the
messenger sectors can have accidental symmetries analogous to the
baryon number giving the particles stability
\cite{gaugemed2,gaugemed1}.  Other natural possibilities include
theories with discrete gauge symmetries \cite{hamaguchi} and string/M
theory \cite{benakli}.

To explore the dynamics which we believe is typical towards the end of
inflation, we primarily focus on two coupled scalar fields in an
expanding universe.  The action can be written as
\be
S = S_g+ S_M
\ee
where
\begin{eqnarray}
S_g & = & -\int d^4x \sqrt{-g} \frac{\mpl^2 R}{16 \pi} \\
S_M & = & \int d^4x \sqrt{-g} \left\{ \frac{1}{2} \left[ g^{\mu \nu}
\phi_{,\mu}
\phi_{,\nu} - m_\phi^2  \phi^2 \right]  + \right. \nonumber \\
&  & \left. \frac{1}{2} \left[ g^{\mu \nu} X_{,\mu}
X_{,\nu} - (m_X^2 + \xi R + g^2 \phi^2) X^2 \right]
\right\}.
\end{eqnarray}
We will take $\xi=1/6$ corresponding to conformal coupling to
gravity although our main results will be insensitive to this
assumption.  Neglecting the small effects coming from the quantum 
fluctuations, we take the gravitational field and the inflaton field
to be purely classical fields.  Note that we are neglecting other
fields which the inflaton field needs to couple to in order to reheat
the universe.  We have numerically verified that as long as the
reheating or preheating occurs on a time scale that is greater
than $5 m_\phi^{-1}$, our main conclusions are insensitive to this
assumption since the particle production that is mainly of interest to
us occurs during this time (as we will later see, nonresonant,
nonperturbative production is of interest for superheavy dark matter
production).

We will consider a metric of the form $ds^2= dt^2- a^2(t) d{\bf x}^2$.
The resulting equations of motion for the homogeneous classical fields
are
\begin{eqnarray}
\frac{\dot{a}^2}{a^2} - \frac{4 \pi}{3} ( \dot{\phi}^2 + m_\phi^2
\phi^2) & = &0 \label{eq:friedmann}\\
\ddot{\phi} + 3\frac{\dot{a}}{a} \dot{\phi} + m_\phi^2 \phi & = &0
\label{eq:phievolve}
\end{eqnarray}
where we have neglected the dark matter contribution to the energy
density. This is a good approximation during the time period of
dynamical interest.

Of course, we do not expect any of our results to be sensitive to the
initial conditions, since our results depend upon what happens towards
the end of inflation and afterwards.  Our results will mainly depend
upon the functional form of the potential for the inflaton field.  In
this paper, we will not study this model dependence but will study
what we consider to be the typical dynamics of such systems.  For
the sake of completeness, we discuss in the Appendix the initial
conditions that we use for our study.

Now let us consider the $X$ sector.  With the canonical conjugate to $X$ as
$a^3 \dot{X}$ and canonically quantizing this action, the Heisenberg
equation of motion is
\be
\ddot{X} + 3 H \dot{X} - \frac{1}{a^2}\nabla^2 X +
(M_X^2 + g^2 \phi^2) X =0
\ee
where $H=\dot{a}/a$ is the Hubble expansion rate.  We introduce
the Fourier convention
\be
X= \int \frac{d^3 k}{(2 \pi)^{3/2} a} (a_k e^{i \vec{k} \cdot \vec{x}}
h_k(t) + a^\dagger_k e^{-i \vec{k} \cdot \vec{x}} h^*_k(t))
\ee
where we have defined $h_k=X_k a$ and defined the normalization for
the annihilation operator as $[a_{\vec{k_1}}, a^\dagger_{\vec{k_2}}]=
\delta^{(3)}(\vec{k_1}-\vec{k_2})$.  Imposing the canonical commutation
condition, we obtain the normalization condition
\be
h_k \dot{h}_k^*-h_k^* \dot{h}_k= \frac{i}{a}.
\ee
The mode equation satisfied by $h_k$ is 
\be
\ddot{h_k} + H \dot{h}_k + \left[ - H^2 - \frac{\ddot{a}}{a} +
\left(\frac{k}{a} \right)^2 + (M_X^2 + g^2 \phi^2) \right] h_k=0.
\label{eq:modeproper}
\ee
In conformal coordinates defined by $ds^2= a^2(\eta) ( d\eta^2 - d{\bf
x}^2)$, this mode equation becomes
\be
h_k''(\eta) + \omk^2 h_k=0
\ee
where $\omk=\sqrt{ k^2 + (M_X^2 + g^2 \phi^2(\eta)) a^2(\eta)}$ and
the $'$ derivative is with respect to conformal time.

Now we need to fix the boundary conditions.  Because the particle number
can be constant only for time translationally invariant systems, the
no-particle state (the vacuum state) existing towards the end of
inflation can be specified only approximately in an expanding
universe (see, for example, \cite{bunch} and \cite{birrelldavies}).   One
method of systematically classifying the various 
inequivalent approximate vacuum states is through the adiabatic vacuum
\cite{bunch} definition.  As will be seen later, we will use an effectively
infinite adiabatic order vacuum boundary conditions by considering the
boundary conditions placed at arbitrarily far past and future for the
nonsingular spacetime that we will consider.

If we denote $h_k^{\eta_1}$ as the mode solution with boundary
conditions defined at a future time $\eta_1$ and $h_k^{\eta_0}$ as the mode
solution with the boundary conditions defined at a time in the past
$\eta_0$, we can define the Bogoliubov transformation as
$h_k^{\eta_1}(\eta)= \alpha_k h_k^{\eta_0}(\eta) + \beta_k h_k^{*
\eta_0}(\eta)$.  The number density then is given by
\be
n_X(t)=\int_0^{\infty} \frac{dk}{2 \pi^2 a^3(t)} k^2 |\beta_k|^2.
\ee

\vspace{36pt}
\centerline{\bf III. STEEPEST DESCENT METHOD}
\vspace{24pt}
This section presents a derivation of the analytic estimation
(\eqr{eq:goodapprox}) that 
will be used in conjunction with the numerical work.  A reader
interested only in its application should skip to the next section.
Its direct application to our physical system of main interest
(presented in Section II) will be given in Section IV. 
The following analysis is in the spirit of Ref.\ \cite{berezin}.

With the definition 
\be 
h_k= \frac{\alpha_k}{\sqrt{2 \omk}} e^{-i \int \omk
d\eta} + \frac{\beta_k}{\sqrt{2 \omk}} e^{i \int \omk d\eta} 
\ee 
the
differential equation 
\be 
h_k'' + \omk^2 h_k =0 
\label{eq:modeconform}
\ee 
is equivalent to
\begin{eqnarray}
\alpha_k' & = & \frac{\omk'}{2 \omk} \exp\left(2 i \int \omk d\eta
\right) \beta_k 
\nonumber \\ 
\beta_k' & = & \frac{\omk'}{2 \omk} \exp\left(-2 i \int \omk d\eta
\right) \alpha_k. 
\end{eqnarray}
Because $\omk'/(2 \omk)$ vanishes at $\eta=\pm \infty$ (adiabatic
in-out region assumption), $\alpha_k$ and $\beta_k$ become constants
there, assuming no singular behavior occurs there.  Expanding
$\alpha_k$ and $\beta_k$ in an adiabatic series (in powers of
derivatives), and using the boundary 
condition $\alpha_k(\eta_p)=1$ and $\beta_k(\eta_p)=0$ (equivalent
to an infinite adiabatic order boundary condition in the limit $\eta_p
\rightarrow -\infty$ for our restricted class of spacetime) we obtain
\be
\beta_k \approx \int d\eta \frac{\omk'}{2 \omk} \exp\left( -2 i \int^\eta
\omk(\eta') d\eta' \right). 
\label{eq:adiab1}
\ee
to leading order.   Note that this approximation should be good as
long as $\beta_k \omk'/(2 \omk) \ll 1$ even when $\omk'/(2 \omk) > 1$.
%
%
%
This is certainly true for the cases to which we wish to apply this analysis.
Our next objective is to obtain an approximation for this integral.
Let us write 
\be
\omk=\sqrt{ k^2 + M_X^2 C(\eta)}
\label{eq:omk}
\ee
where all the $\eta$
dependence is contained in $C(\eta)$ and the radicand is positive
definite for all real $\eta$.  For example, in our model,
\be
C(\eta)= \left( 1+ \frac{g^2 \phi^2(\eta)}{M_X^2} \right) a^2(\eta)
\label{eq:ceffourmodel}
\ee
which can be thought of as the square of an ``effective'' scale
factor.  We will also assume that $C(\eta)$ is $C^\infty$ for real
$\eta$ and analytically continue $\omk$ to the complex plane.
Because of the squareroot in the exponent, the poles of the integrand
in \eqr{eq:adiab1} will also be branch points in the complex $\eta$
plane.\footnote{We assume that 
the integral of the squareroot in the exponent will introduce no other
branch points.}   We will choose the branch cuts such that they go
from the branch points to infinity along a path such that the branch
points on the lower half plane  have the cut
going towards $- i \infty$ and the branch points on the upper half
plane have the cut going towards $i \infty$.  Furthermore, the cut
will be taken along the curve where the exponential function has an
equal modulus.  Transverse to the cut, the exponential will fall off
rapidly.

\begin{figure}
\end{figure}
\begin{figure}[t]
\hspace*{25pt} \epsfxsize=400pt \epsfbox{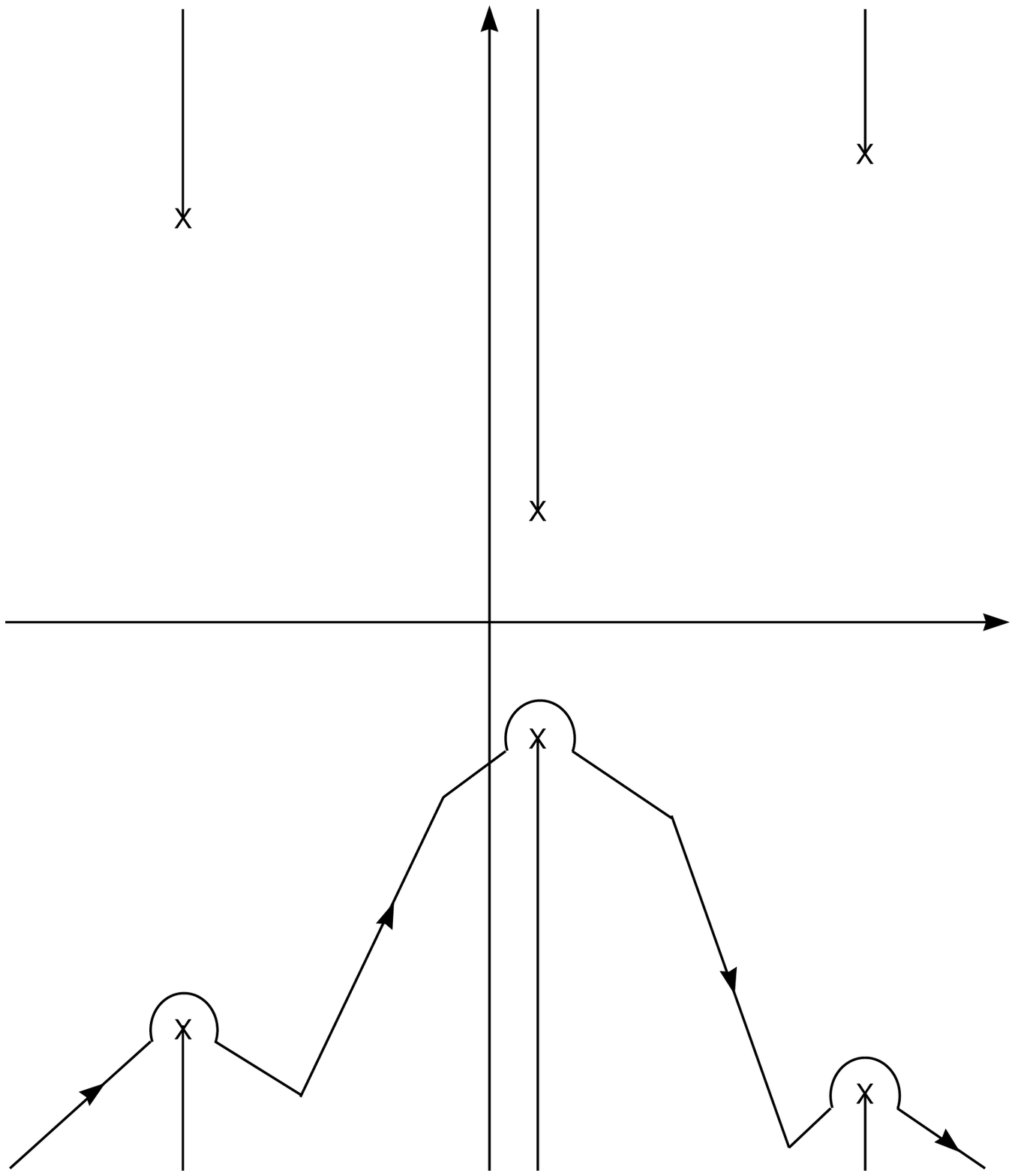}
\caption{ A schematic sketch of the analytic structure of
\eqr{eq:adiab1} on the complex $\eta$ plane is shown.  The crosses
represent branch points and the lines emanating from them  branch
cuts.  Shown also is a schematic sketch of the appropriately deformed
contour for the steepest descent approximation on the lower half plane. 
}
\label{fig:contour}
\end{figure}

The integral over the real axis can be replaced (using Cauchy's
theorem) with the integral over an appropriately deformed  
contour in the lower or upper half plane (we will soon see that our
phase convention is such that we are really concerned with the poles
on the lower half plane as shown in Fig.\ \ref{fig:contour}).  The main 
contribution from the integral over the deformed contour will come
from near the 
branch points and possibly the end points.  The branch points will be
distributed symmetrically with respect to reflection across the real
axes because of the Schwartz reflection principle.  The end point contribution
will be of the order of $\omk'/\omk$.  However, in our restricted class
of spacetimes which admits an infinite adiabatic order vacua, we can
make this contribution arbitrarily small by taking the endpoints
further out.  (We comment further on this effect later.)  Hence in
general, \eqr{eq:adiab1} can be approximated as 
a coherent sum of steepest descent integrals around each of the branch
points.

Let us look at the contribution from the $j$th branch point denoted as
$\ett_j$.  Near this branch point, the integral in
the exponent of \eqr{eq:adiab1} can be expanded as
\be
\int_{\eta_p}^\eta \omk(\eta) d\eta =  \int_{\eta_p}^{\ett_j}
\omk(\eta) 
d\eta + \frac{2 M_X}{3} \sqrt{C'(\ett_j)} \delta^{3/2} + . . .
\ee  
where $\delta \equiv \eta-\ett_j$ and we kept the leading term in the
$\delta$ expansion of $C$ (we will assume that $C'$ does not vanish
here).  Expanding in a similar way 
$\omk'/\omk$ in \eqr{eq:adiab1}, we can write the contribution to
$\beta_k$ from this branch point as
\be
U_j \equiv v_j \exp(-2 i \int_{\eta_p}^{\ett_j} \omk(\eta) d\eta)
\label{eq:uj}
\ee
where we defined $v_j$ as
\be
v_j \equiv \frac{1}{4} \int_{C_j} \frac{d \delta}{\delta}
\exp(\frac{-4 i}{3} M_X
\sqrt{C'(\ett)} \delta^{3/2} ) 
\label{eq:contourint}
\ee 
and $U_j$ was introduced to rewrite $\beta_k$ as
\be
\beta_k \approx \sum_j U_j.
\label{eq:sumit}
\ee
Here the contour $C_j$ near $\ett_j$ is the steepest descent contour.

Let us determine the steepest descent path near $\ett$.  If we denote
$\theta_2 \equiv \arg(C'(\ett))$ and denote $\theta$ to be the
argument of $\delta$ along the steepest descent contour, the
restriction on $\theta$ is 
\be 
\theta = \frac{\pi (4 n -1) - \theta_2}{3}
\label{eq:steepcond}
\ee
where $n$ is an integer.  Now, if we let the branch cut go along
$\arg(\delta)=\alpha$ such that 
$\arg(\delta) \in [\alpha, \alpha+2 \pi)$ on the lower half plane and
$\arg(\delta) \in (\alpha-2 \pi, \alpha]$ on the upper half plane, upon
choosing $\sqrt{C'(\eta)} > 0$ when $\eta$ is real (consistent
with positive frequency mode definition), we can place
the restriction $ \theta_2 \in [-\alpha,  2 \pi -\alpha)$ in the upper
half plane and $ \theta_2 \in (-2 \pi - \alpha, - \alpha]$ in the lower
half plane.  As mentioned above, we choose the branch cuts to go towards
$\pm i \infty$ by restricting $\alpha \in (0, \pi)$ on the upper half
plane and $\alpha \in (-\pi,0)$ on the lower half plane.  Finally,
restricting the branch cut to be  on an equal modulus
curve, we obtain the relationship
\be
\alpha = \frac{-\theta_2 \pm 2 \pi }{3} 
\label{eq:branchcut}
\ee
where the positive sign corresponds to the upper half plane and the
negative sign corresponds to the lower half plane.

Combining Eqs.\ (\ref{eq:steepcond}) and (\ref{eq:branchcut}), we find
two possible integers $n=0,1$ for the lower half plane, while we find only
one possible integer $n=0$ for the upper half plane.  Since the steepest
descent approximation requires an incoming direction and an outgoing
direction for each of our branch points, we can only use the lower
half plane branch points.  To summarize, the necessity of making
branch cuts with the appropriate shape and the information regarding
the derivative of $C$ at the branch point determines the usable
steepest descent paths, which are restricted to the lower half plane
(lower and not upper because of the sign convention on the
squareroot).  This, as we will see below, corresponds to an
exponentially damped result.

Let us return to the evaluation of $v_j$ in \eqr{eq:contourint}.  
From \eqr{eq:steepcond} and \eqr{eq:branchcut}, we see that the branch
cut bisects the angle made by the steepest descent contour which makes
a fixed (acute) angle of $2 \pi/3$ (or $4 \pi/3$ in coordinate angle),
and hence the steepest descent 
path is well defined independently of $\theta_2$.  Hence, we can
easily evaluate the integral after making a change of variables
$u=\delta^{3/2} (-4i m\sqrt{C'(\ett_j)}/3)$.  The result is
\be
v_j= \frac{i \pi}{3}
\label{eq:vjsol}
\ee

Now we need to evaluate
\be
\int_{\eta_p}^{\ett_j} \omk(\eta) d\eta
\label{eq:mainintegral}
\ee
to complete our evaluation of $U_j$.  Since we do not in general know
the function $C(\eta)$ on the complex plane but know it and
its derivatives on the real axis (numerically for the model presented
in section II), let us
find an estimation utilizing that $\omk$ is analytic around the real
axis.  Let $\ett_j=r+\mu$ where $r$ is purely real and
$\mu$ is purely imaginary.  Now, we split the integral into
\be
\int_{\eta_p}^{\ett_j} \omk(\eta) d\eta = \Phi_j  + J_j
\ee
where
\begin{eqnarray}
\Phi_j &=& \int_{\eta_p}^{r} \omk(\eta) d\eta  \\
J_j &=&\int_{r}^{\ett_j} \omk(\eta) d\eta 
\end{eqnarray}
where $\Phi_j$ is purely real.  To evaluate $J_j$, we expand in Taylor series,
\be
J_j \approx \omk(r) \mu + \omk'(r) \mu^2/2 + \omk''(r) \mu^3/6 + . . .
\ee
where one should note that all the even terms are real.  Because we
will mainly be interested in one pole domination case, we will only
calculate the imaginary value of $J_j$.  Thus, as long 
as
\be
|\omk''/ \omk| \ll |6/\mu^2|
\label{eq:leadingwjust}
\ee
we can truncate the $J_j$ after the first term.  We shall check the
self consistency later.

To approximate $\ett_j$, we Taylor expand the left hand side of $C=
-(k/M_X)^2$ about $r$ to obtain
\begin{eqnarray}
\frac{\mu^2}{6} C'''(r) + C'(r) & = & 0 \label{eq:realpart} \\
\frac{\mu^2}{2} C''(r) & = & - \frac{\omk^2(r)}{M_X^2} \label{eq:musolve}
\end{eqnarray}
where $C''(r) >0$ since $\mu$ is purely imaginary.
The truncation of the Taylor expansion should be justified as long as 
\begin{eqnarray}
\left| \frac{C^{(5)}(r)}{C'''(r)} \right| \frac{ \omk^2}{10 M_X^2
C''(r)} & \ll  & 1 
\label{eq:rvalid} \\
\frac{|C^{(4)}(r)|}{C''(r)^2}\frac{ \omk^2}{6 M_X^2}  & \ll & 1
\label{eq:muvalid}
\end{eqnarray}
for Eqs.\ (\ref{eq:realpart}) and (\ref{eq:musolve}) respectively where
the superscript indicates the order of the derivative.\footnote{Note that even
if \eqr{eq:rvalid} is not satisfied, all the equations derived may
still be valid as long as $r$ can be estimated another way.}

Assuming that the contribution from one of the poles dominates in
\eqr{eq:sumit}, we can approximate $|\beta_k|^2 \approx |U_1|^2$.
Using \eqr{eq:musolve} and  choosing the solution on the lower half
plane (as we have justified above), we obtain
\be
|\beta_k|^2 \approx \exp\left( -4 \left[ \frac{(k/\aeff)^2}{M_X
\sqrt{\heffsq + \reff/6}} + 
\frac{M_X}{\sqrt{\heffsq+\reff/6}} \right] \right)
\label{eq:goodapprox}
\ee
where we have dropped the factor of $(\pi/3)^2$, we have defined the
effective scale factor 
\be
a_{\mbox{eff}} = \sqrt{C},
\label{eq:defaeff}
\ee
and $\heff$ and $\reff$ correspond to Hubble expansion rate and Ricci scalar
(respectively) for the metric $ds^2= a_{\mbox{eff}}^2(\eta) ( d\eta^2 - d{\bf
x}^2)$.  All time dependent 
quantities in \eqr{eq:goodapprox} are evaluated at $\eta=r$ given by
\eqr{eq:realpart}.  Explicitly, the radicand of the exponent in
\eqr{eq:goodapprox} is simply $\heffsq + \reff/6=C''(r)/(2 C^2(r))$
since $6 a''_{\mbox{eff}}/a^3_{\mbox{eff}} = \reff$ and 
$a'_{\mbox{eff}}/a^2_{\mbox{eff}}= \heff$.  Rewriting the condition
\eqr{eq:leadingwjust} 
in this new notation as  (in the $k \rightarrow 0$ limit)
\be
6 \heffsq + \reff \gg \reff/6
\ee
we see that this  condition is almost always satisfied as long as the 
\eqr{eq:muvalid} is satisfied.\footnote{\eqr{eq:muvalid} must be satisfied
since we are using \eqr{eq:musolve} to get the value of $\mu$.}

The exponentially suppressed behavior is not exact for arbitrarily
large masses since the endpoint contribution will eventually become
larger.  An asymptotic analysis of the problem in the appendix of
\cite{ckr1} gives the general power law mass dependence of the endpoint
contribution.  As discussed there, this contribution can become
particularly significant when $C$ is not analytic.  The present
analysis, however, shows that owing to the fact the endpoints lie far
away from the branch points (which have stationary phases), in the
case that $C$ is analytic, the contribution from the endpoints can be
made arbitrarily small as long as the spacetime admits an infinite
adiabatic order in-out regions.

Let us restate the main point of this somewhat technical section.
Given a homogeneous classical field coupling that gives rise to mode
equations of the form \eqr{eq:modeconform} with $\omk$ given by
\eqr{eq:omk} and given an in-out infinite adiabatic order vacua, the
Bogoliubov coefficient giving the particle density per mode is given
by \eqr{eq:goodapprox} and the number density is given by\footnote{
Although $r$ in general depends on $k$ (for an explicit example, see
\eqr{eq:tanhsol} ), because most of the 
contribution to the number density comes from 
$(k/M_X)^2 \ll 1$ where the $k$ dependence can be neglected, we shall
neglect $k$ when doing the momentum integral.} 
\be
n_X(t) \approx \frac{a^3_{\mbox{eff}}(r)}{8 \pi^{3/2} a^3(t)} \exp\left(
\frac{-4 M_X}{\sqrt{\heffsq + \reff/6}}\right) \left[\frac{M_X}{4}
\sqrt{\heffsq 
+\reff/6}\right]^{3/2} 
\label{eq:numdensity}
\ee
where $\aeff$, $\reff$, and $\heff$ are effective scale factor,
Ricci scalar, and Hubble expansion rate defined by \eqr{eq:defaeff}
and the statement following it.  The ``effective'' quantities are all
evaluated at a value of $r$ satisfying Eqs. (\ref{eq:realpart}) and
(\ref{eq:musolve}).  As will be discussed towards the end of next
section, $r$ is roughly the point at which $C$ varies
nonadiabatically, or a bit more precisely, where $C''(r)/C^2(r)$ is a
maximum.  For couplings of the form $Q(\phi) X^2/2 + \xi R X^2/2$
where $Q$ is analytic and $R$ is the Ricci scalar of the spacetime, we
can write
\be
C(\eta)= a^2(\eta) ( 1 + Q(\phi)/M_X^2 + (\xi -1/6)R/M_X^2)
\label{eq:example}
\ee
in a spacetime with the metric $ds^2= a^2(\eta) (d\eta^2 - d{\bf x}^2)$.


\vspace{36pt}
\centerline{\bf IV. COMPARISON WITH EXACT RESULTS}
\vspace{24pt}

Let us first apply \eqr{eq:goodapprox} to a couple of exactly soluble
cases.  Consider the case given by Ref.\ \cite{audretsch} when there
is no inflaton coupling to the $X$ field and the spacetime is given by
$ds^2= C(\eta) (d\eta^2 - d{\bf x}^2)$ where
\be
C(\eta)=  c_1^2 + c_2^2 \eta^2
\ee
where $c_1$ and $c_2$ are real constants.  The exact number density per
mode is
\be
|\beta_k|^2 = \exp\left(- \pi \left[\frac{k^2}{M_X  c_2} + \frac{M_X
c_1^2}{c_2} \right] \right)
\label{eq:exact1}
\ee
which has been noted \cite{audretsch} as the spectrum of a
nonrelativistic gas of particles with momentum $k/\sqrt{C}$ having
chemical potential of $- M_X c_1^2/(2 C)$ and a temperature of $c_2/(2
\pi C)$ 

In deriving \eqr{eq:goodapprox}, we have made two separate
approximations.  One is the usage 
of the steepest descent method and the other is the estimation of the
integral \eqr{eq:mainintegral}.  To test the goodness of each of these
approximations, 
since we know $C$ exactly here, we will first consider the steepest
descent method with the exact branch point.  Let's start from Eqs.\ 
(\ref{eq:uj}) and (\ref{eq:sumit}).  One branch point is on the lower
half plane at  
\be
\eta= \ett \equiv - i \sqrt{\frac{c_1^2 + k^2/M_X^2}{c_2^2}}
\label{eq:branchptexact}
\ee
and another, its complex conjugate, is on the upper half plane.  The
quantity $v_j$ has been already evaluated in general, corresponding to
a contour integral around the branch point on the lower half plane,
and is given by \eqr{eq:vjsol}.  The integral \eqr{eq:mainintegral}
can be easily calculated with $\ett_j=\ett$ to give
\be
\int_{\eta_p}^{\ett_j} \omk(\eta) d\eta = ( -i \pi/2 + \mbox{real
phase}).
\ee
Putting this in \eqr{eq:uj}, we have
\be
|\beta_k|^2 \approx \left(\frac{\pi}{3}\right)^2 \exp\left(- \pi
\left[\frac{k^2}{M_X c_2} + \frac{M_X c_1^2}{c_2} \right] \right)
\label{eq:partialapprox}
\ee
whose exponent we see matches the exact result of \eqr{eq:exact1}.

Now, let's apply \eqr{eq:goodapprox}.  \eqr{eq:realpart} gives $r=0$.
Hence, we find
\be
|\beta_k|^2 \approx \exp\left(- 4 \left[\frac{k^2}{M_X
c_2} + \frac{M_X c_1^2}{c_2} \right] \right)
\label{eq:fullapprox}
\ee
which gives only an approximately correct exponent.  However, note
that the functional form of the mass dependence of the exponent is
exact.  Hence, although the steepest descent approximation seems to give
accurate results for the exponent, as exemplified by
\eqr{eq:partialapprox}, the Taylor expansion method used to estimate the
integral \eqr{eq:mainintegral} leads to a  numerical value of the
exponent that is only roughly correct as seen in \eqr{eq:fullapprox}.
Still we see the functional 
behavior of the mass dependence is accurate.

Let's consider another exactly soluble case \cite{bernard} (see also
\cite{birrelldavies}) specified by
\be
C(\eta) = A + B \tanh(\rho \eta)
\label{eq:tanhmodel}
\ee
where $A$, $B$, and $\rho$ are positive constants with $A > B$,
This results in particle density per mode of
\be
|\beta_k|^2 = \frac{ \sinh^2\left\{\frac{\pi}{2 \rho} \left[ \sqrt{k^2
+ M_X^2 (A 
+ B)} -\sqrt{k^2 + M_X^2 (A-B)}\right] \right\} }{ \sinh\left[
\frac{\pi}{\rho} \sqrt{ 
k^2 + M_X^2 (A-B) } \right] \sinh\left[ \frac{\pi}{\rho} \sqrt{k^2 +
M_X^2 (A+B) 
} \right] } .
\ee
When the exponential cut off starts to become effective for large
masses, the behavior is
\be
|\beta_k|^2 \approx  \exp\left( -\frac{2 \pi M_X}{\rho} \sqrt{ A-B}
\right)
\label{eq:tanhlargem}
\ee 
where we have effectively set $k=0$.  Let us compare this with the
steepest descent approximation.  In this example, it is easier to find
the branch point exactly.  However, since we are interested in using
the Taylor expansion estimation, let us solve Eqs. (\ref{eq:realpart}) and
(\ref{eq:musolve}) which are valid if the conditions
Eqs.\ (\ref{eq:rvalid}) and (\ref{eq:muvalid}) can be satisfied.  These
conditions are respectively
\begin{eqnarray}
\left| \frac{4 \rho^2 (2 - 17 t_r^2 + 30 t_r^4 - 15 t_r^6)}{(1- t_r^2)(1-3
t_r^2)} \right| &
\ll & \frac{-20 B \rho^2 (1- t_r^2) t_r}{(k/M_X)^2 + A + B t_r} \\
\left| \frac{4 \rho^2 t_r ( 2 - 5 t_r^2 + 3 t_r^4)}{(1- t_r^2) t_r} \right| &
\ll & \frac{-12 B 
\rho^2 (1-t_r^2) t_r}{(k/M_X)^2 + A + B t_r}
\end{eqnarray}
where we have defined $t_r \equiv \tanh(\rho r)$.  Here, we know that $t_r
<0$ because 
\be
C''(r) = -2 B \rho^2 (1- t_r^2) t_r
\ee
must be positive as we explained before.  Hence, we only need to consider
$t_r \in (-1, 0]$ and find that our conditions imply $A \approx B$.
Solving Eqs. (\ref{eq:realpart}) and (\ref{eq:musolve}) for $t_r$, we
find
\be
t_r=\frac{-B - \sqrt{B^2 + 3 (A+(k/M_X)^2)^2}}{3 (A+(k/M_X)^2)}
\label{eq:tanhsol}
\ee
which then implies
\be
\frac{C''(r)}{2 C^2(r)} = \frac{4 \rho^2}{9 (A-B)} - \frac{\rho^2}{2 B}
+ {\cal O}(A-B)
\ee
where we have expanded in powers of $A-B$ and set $k=0$.  Therefore,
in the large mass limit, using again \eqr{eq:goodapprox}, we have
\be
|\beta_k|^2 \approx  \exp\left( -\frac{6 M_X}{\rho} \sqrt{ A-B}
\right)
\label{eq:tanhtaylorapprox}
\ee
which is in reasonable agreement with \eqr{eq:tanhlargem}.  The lesson
that we have learned here is that as long as one accounts for the
regime in which the approximations are valid, the parameter dependence
of the damping exponent is accurately calculated by our method.

Finally, we note that an intuitive meaning can be attached to the
value of $r$.  The quantity $H^2_{\mbox{eff}}(\eta) +
R_{\mbox{eff}}(\eta)/6$ has a maximum 
 at an $\eta=\eta_*$ satisfying the equation
\be
C' C'' + C C'''=3 C' C''
\label{eq:curvmax}
\ee
which\footnote{Taking into account the validity condition for the
Taylor expansion,  it is easy to see that $\eta_*$ indeed
corresponds to a maximum.} is very similar to the equation determining $r$
\be
(k/M_X)^2C''' + C C''' = 3 C' C''.
\ee
Since $(k/M_X)^2 C''' \sim C' C''$ (as can be seen by looking at
Eqs. (\ref{eq:curvmax}) and (\ref{eq:goodapprox})) for important
values of $k$ and since $\eta_*$ is a stationary point, we
expect $r$ to be very close to $\eta_*$.  Therefore, $r$ in general
will roughly correspond to the most ``nonadiabatic'' point, i.e. the
point at which $H^2_{\mbox{eff}}(\eta) + R_{\mbox{eff}}(\eta)/6$ is a
maximum. 

For example, let us consider the model specified by
\eqr{eq:tanhmodel}.  If one takes the branch point determining
equation to be \eqr{eq:curvmax}, then the resulting value of the
number density per mode is
\be
|\beta_k|^2 \approx  \exp\left( -\frac{4 \sqrt{2} M_X}{\rho} \sqrt{ A-B}
\right)
\ee
which is in reasonable agreement with
Eqs. (\ref{eq:tanhtaylorapprox}) and (\ref{eq:tanhlargem}).


\vspace{36pt}
\centerline{\bf V. ANALYTIC ESTIMATION}
\vspace{24pt}
Let us first consider the work of Ref.\ \cite{ivaylo} to see what part
of the parameter space we are interested in.  Ref.\ \cite{ivaylo}
estimated the particle production in our type of system by
neglecting any contribution from the decaying mode and using the
Mathieu instability band plot to estimate the ``average'' exponent of
the mode growth.  (Recall that in an expanding universe, the mode
equation is not exactly the Mathieu equation, since the parameters of
what would be the Mathieu equation is time dependent in an expanding
universe.)  This method gives a good idea of when the particle
production will be efficient through the resonance phenomenon.


Let's start with the number density per mode written as
\be
|\beta_k|^2 = \frac{\Omega_k}{2} \left( | \chi_k|^2 + \frac{1}{\Omega_k^2}
|\dot{\chi_k} - \frac{\dot{a}}{2 a} \chi_k |^2 \right) -1/2
\ee
where $\chi_k$ is the solution to the conformally coupled mode
equation (\ref{eq:modeproper}) with $\chi_k= \sqrt{a} h_k$ satisfying
the $0$th order adiabatic vacuum boundary condition 
in the past and $\Omega_k= \sqrt{(k/a)^2 + M_X^2 + g^2 \phi^2}$.
Approximate the relevant solution as  
\be
\chi_k \sim \frac{\exp(\int
\mu m_\phi dt -i \int w_k dt)}{\sqrt{2 w_k}}
\ee
 where $w_k=\sqrt{\Omega_k^2 +
(1/4) H^2 - (1/2) \dot{H}}$ and $\mu$ is the Mathieu
characteristic exponent when the mode equation is written in the form
of \eqr{eq:mathieu} (see below).  The form of the approximation is
roughly equal to the lowest order adiabatic approximation and taking
only the growing mode is justified by the fact that the decaying
solution will be small in comparison.  Now, approximating $w_k \approx
\Omega_k$ and assuming $(\mu m_\phi)^2/(2 \Omega_k^2) \ll 1$, we obtain
the approximation
\be
|\beta_k|^2 \sim \frac{1}{2} [ \exp(2 \int \mu m_\phi dt) -1 ]
\label{eq:ivaylostart}
\ee
which is the starting point in \cite{ivaylo}.\footnote{They had used
minimal coupling to gravity, but as we can see that is of little
importance as far as the approximation \eqr{eq:ivaylostart} is concerned.}

The mode equation (\ref{eq:modeproper}) with $\chi_k= \sqrt{a} h_k$,
when written in the form of the Mathieu equation \eqr{eq:mathieu}
gives
\begin{eqnarray}
A &  \approx &  \left( \frac{k}{a m_\phi} \right)^2 + 
\left( \frac{M_X}{m_\phi} \right)^2 + 2 q + 
2/(3 m_\phi t)^2 \nonumber \\
q & \approx & \frac{g^2 \Phi(t_0)^2}{4 m_\phi^2 (m_\phi t)^2}
\label{eq:aqparams} 
\end{eqnarray}
where we have chosen the initial time to be at $t=t_0=1/m_\phi$ and
$\Phi(t_0)$ represents the amplitude of the inflaton field
oscillations at the initial time when the inflaton field oscillation
amplitude starts to decay like $1/t$ in a pressureless universe.
Ref. \protect{\cite{ivaylo}} then considers the trajectory of these parameters
as a function of time and estimates the exponent integral in
\eqr{eq:ivaylostart}.  They find that the efficient preheating ceases
at around $M_X=100 m_\phi$ for $g < 1$.  After that, the incoherent
decay process dominates the energy release of the inflaton.

\begin{figure}[t]
\hspace*{25pt} \epsfxsize=400pt \epsfbox{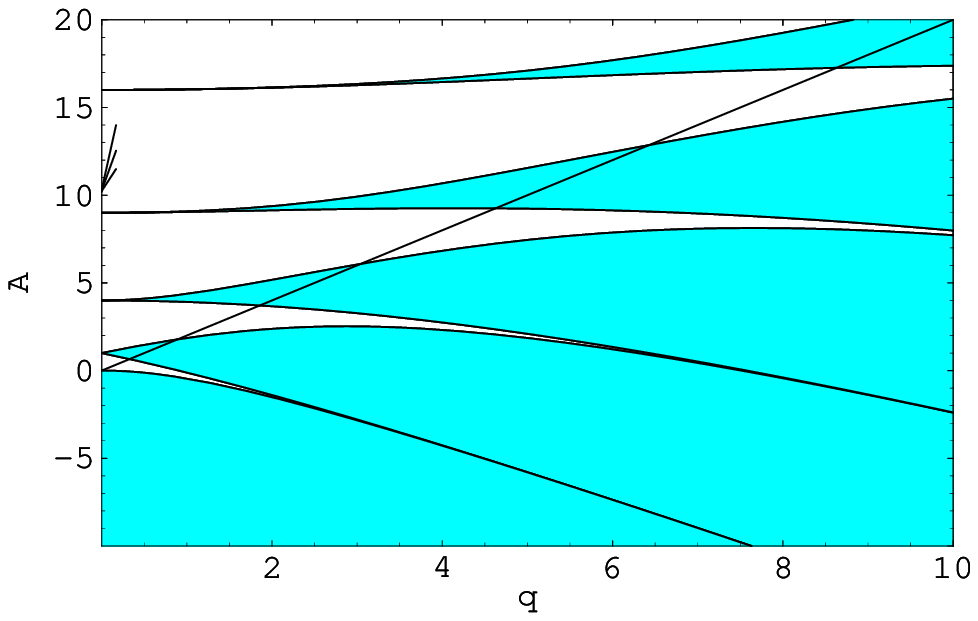}
\caption{Shown is the Mathieu instability chart.  For the values of
$A$ and $q$ that fall in the shaded region, the solution to the
Mathieu equation is unstable.  The line connecting the origin to the
upper right corner of the plot is $A=2 q$.  All $A$ and $q$
trajectories must lie above this line by their definition in
\eqr{eq:aqparams}.  The three short line segments  near $(q,
A)=(0,11)$ represents three representative trajectories each having
different values of $k$ (for those $k$ values that dominate the number
density integral) for $\gparam=45$ and $b=7$.  Although none of the
trajectories cross any of the instability bands, there is an
enhancement by a factor 100 in the particle density production.}
\label{fig:notenough}
\end{figure}

In our case, we are more interested in the regime in which the
resonance phenomenon is extremely ineffective, or virtually
non-existent, because only an extremely tiny density of very massive
particles is allowed for the dark matter scenario.   Define the
parameter $b=M_X/H_i$ where $H_i$ is the value of the Hubble parameter
at the end of inflation.  Consider the case when $b=7$ and $g=0$, for
which $\omhsq=0.128$.
When the interaction is turned on to say
$\gparam = 45$, we find that  there is a factor of 100 increase in the
particle density produced to $\omhsq=12$.  As illustrated in Fig.\
\ref{fig:notenough}, we see that for this value of $\gparam$, $b$, and
values of $k$ that dominate the integral for the mode sum, the Mathieu
parameter trajectories never cross any of the instability bands.
Thus, the analysis of Ref.\ \cite{ivaylo} is only partially applicable
to our case, mostly because we are only requiring a very small fraction of
the inflaton energy density to turn into $X$ particles.  This means
that the $X$ particle masses that can be produced in 
interesting quantities should be significantly greater than $100
m_\phi$ for coupling constants still less than 1.  Furthermore, since
significant dark matter production occurred without the trajectories ever
crossing any of the instability bands, we can expect sufficient dark
matter to be generated without any resonance effects. 

\begin{figure}[t]
\hspace*{25pt} \epsfxsize=400pt \epsfbox{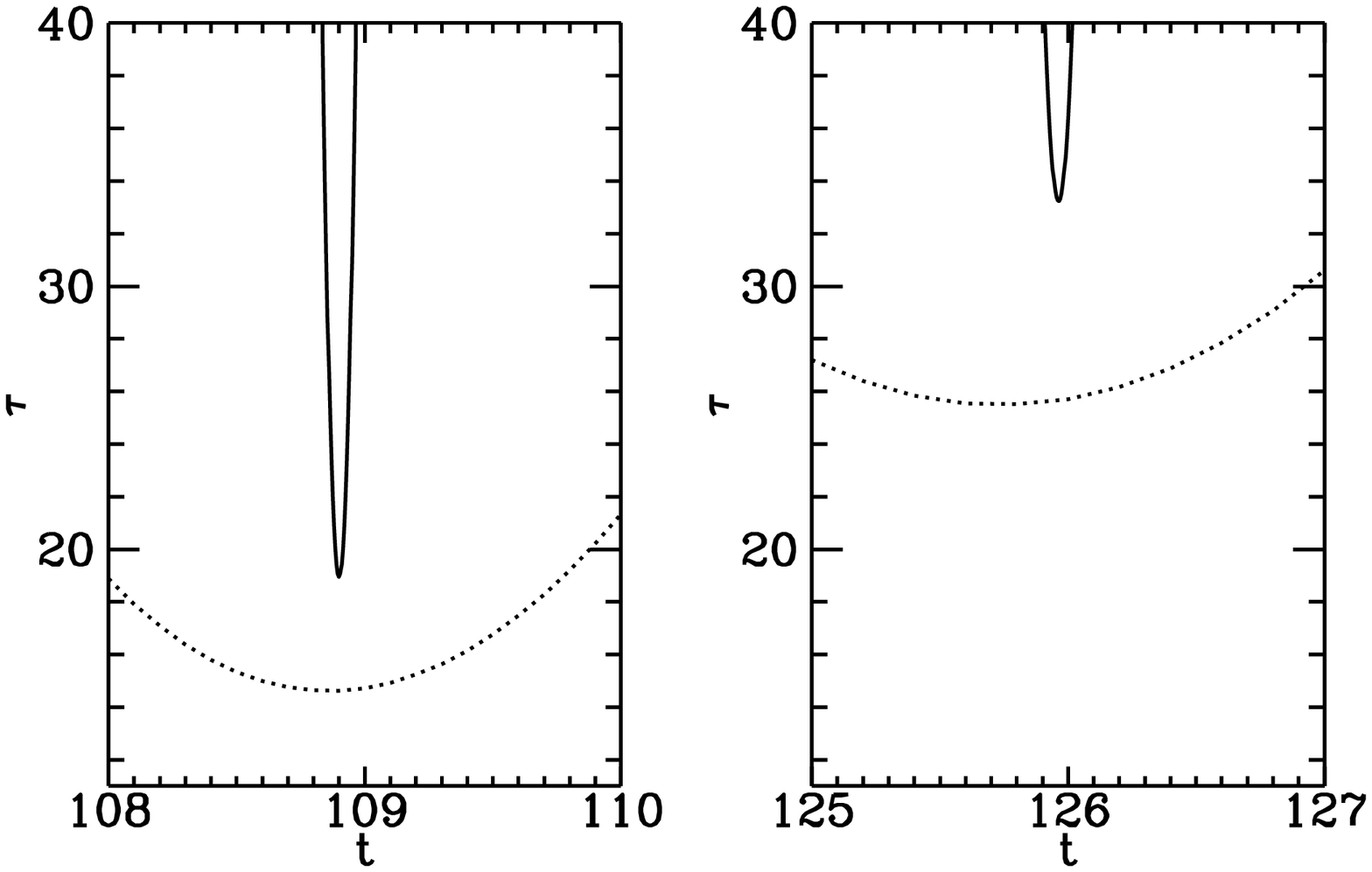}
\caption{Here 
we plot the number density suppressing exponent $\tau
\protect{\equiv} 4 b /(H^2_{\mbox{eff}}(t) +
R_{\mbox{eff}}(t)/6)^{1/2}$ as a 
function of $t$ (in units of $\sim 1/(12.2 H_i)$) 
near the values corresponding to the conformal time $r$ which solves
 Eqs.\ (\protect{\ref{eq:realpart}}) and  (\protect{\ref{eq:musolve}}). 
The solid curve corresponds to the case
$b=1080$ and $\gparam=10^6$.  The dotted curve corresponds to $b=30$ and
$g=10^3$.  The actual values of  $r$ solving Eqs.\
(\protect{\ref{eq:realpart}}) and  (\protect{\ref{eq:musolve}}) with  
these parameters lie very close to the minimum of 
these curves (there exists only one solution $r$ per curve shown).
Exact numerical value examples are given in the text.  The $t(r)$ near
109 and near 126 gives the smallest two $\tau$ values for each
parameter set.  Clearly the pole having a real part $r$ with $t(r)$
near 109 dominates.  }
\label{fig:leadtaus}
\end{figure}

Now let us apply the estimation of \eqr{eq:goodapprox} to the system
presented in Section II.  Let us consider the largest possible
perturbative coupling of around $\gparam = 10^6$ which is what would
give the largest possible mass for the dark matter produced.\footnote{
We are assuming here that something like SUSY is protecting the
inflaton potential from large radiative corrections that can spoil
inflation.}  We first look for the solutions to
Eqs. (\ref{eq:realpart}) and (\ref{eq:musolve}). There are many
solutions to these equations, as we expect.  However, only one of
these solutions will be relevant for the one pole domination
approximation.  To see which solution will be relevant, we plot in
Fig. \ref{fig:leadtaus} the absolute value of the exponent in
\eqr{eq:numdensity} which we will denote as $\tau$, as a function of 
time values that are near the actual 
solution to Eqs.\ (\ref{eq:realpart}) and (\ref{eq:musolve}).  For
definiteness, we will fix $b$ at $1080$ although the one pole
domination approximation will be reasonable for all other masses
within the range of our interest.  It is clear from these plots that
the pole having a real part corresponding to $t \approx 109$ will
dominate in the sum  \eqr{eq:sumit}, thereby justifying the one pole
approximation 
reflected in \eqr{eq:goodapprox}.  In Fig. \ref{fig:leadtaus}, we have
also plotted the case of $\gparam=10^3$ and $b=30$ to indicate the
generic nature of this one pole domination.

The solutions to \eqr{eq:realpart} and \eqr{eq:musolve} that is of
interest to us are not sensitive to the value of $b \equiv M_X/H_i$ as
long $\gparam/b \gg 1$ which turns 
out to be within our range of interest mainly because $\phi< \mpl$ and
$|\dot{\phi}/H_i| < \mpl$ at the end of inflation as can be seen by
looking at the exponent of \eqr{eq:goodapprox}.  For example, with
$k=0$, the value of $t(r)$ that will be of interest to us is $108.90$ (in
units of $1/(12.2 H_i)$) for
both $b=900$ and $b=1800$.  Only for values of $b$ as large as $10^4$
(which we will not be of much interest to us), does the value of $t(r)$
deviate to $108.89$.  The dependence on $k$, which can be seen
directly in Eqs.\ (\ref{eq:realpart}) and (\ref{eq:musolve}) will also
be negligible because the most of the SDM that are produced will be
nonrelativistic.

\begin{figure}[t]
\hspace*{25pt} \epsfxsize=400pt \epsfbox{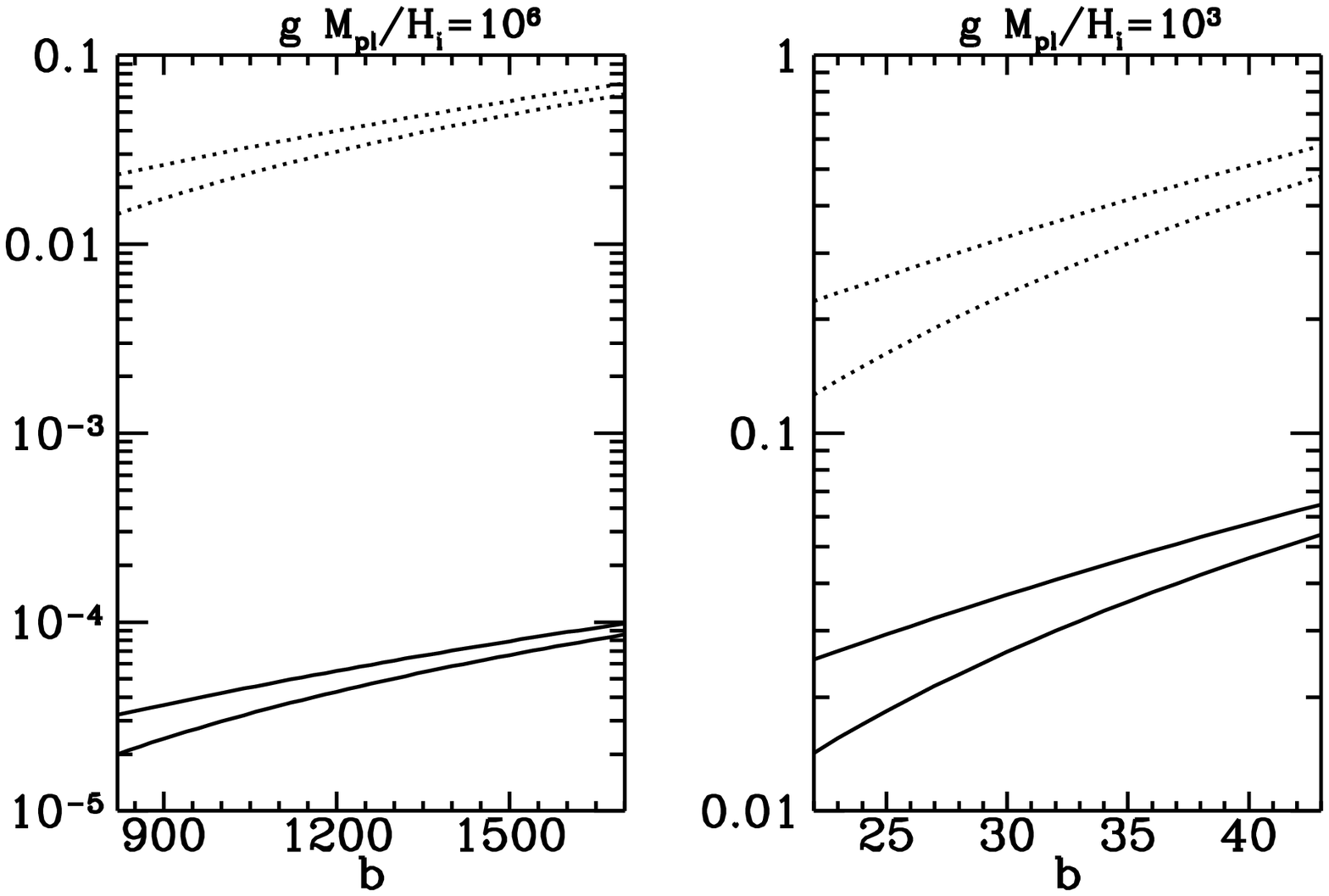}
\caption{  We plot as a function of $b\equiv M_X/H_i$ the quantities
that must be less than unity for the Taylor series truncation to be
valid.  The solid 
curves correspond to the left hand side 
of \eqr{eq:muvalid}, and the dotted curves correspond to the
left hand side of \eqr{eq:rvalid}.  The lower of the given curve type
corresponds to $k/(a_i H_i)=0$.  The upper of the
given curve type corresponds to $k/(a_i H_i) = 1000$ for the
$\gparam=10^6$ case and $k/(a_i H_i) = 30$ for the $\gparam=10^3$
case.  These $k$ values indicate the momentum range over which the
particles are produced.
The expansion truncation is clearly justified in the $\gparam=10^6$
case while it is marginally adequate in the $\gparam=10^3$ case.}
\label{fig:validity}
\end{figure}

Hence, with $t(r)=108.9$ we use \eqr{eq:goodapprox} to find that
$\omhsq=1$ at $b \approx 1450$ and 
$\omhsq=0.01$ at $b \approx 1550$.  Hence, taking $H_i \approx
m_\phi/2$, the maximum possible mass of SDM for
which its abundance will be cosmologically significant is
an order of magnitude above the maximum mass that can accommodate
efficient preheating.  In the next section we will give a bit better
estimate by taking into account the numerical calculation of the
particle production.

Now, let us check the validity of truncating the Taylor expansion in
approximating the location of the poles.   The quantities on the left
hand side of \eqr{eq:muvalid} and \eqr{eq:rvalid} are plotted in Fig.\
\ref{fig:validity}.  From the figure it is clear that for the
$\gparam=10^6$ the approximation is well justified.  However, the
right hand panel of the figure shows that the approximation is only
marginally adequate for $\gparam=10^3$.  

\vspace{36pt}
\centerline{\bf VI. NUMERICAL ESTIMATION}
\vspace{24pt}

\begin{figure}[t]
\hspace*{25pt} \epsfxsize=400pt \epsfbox{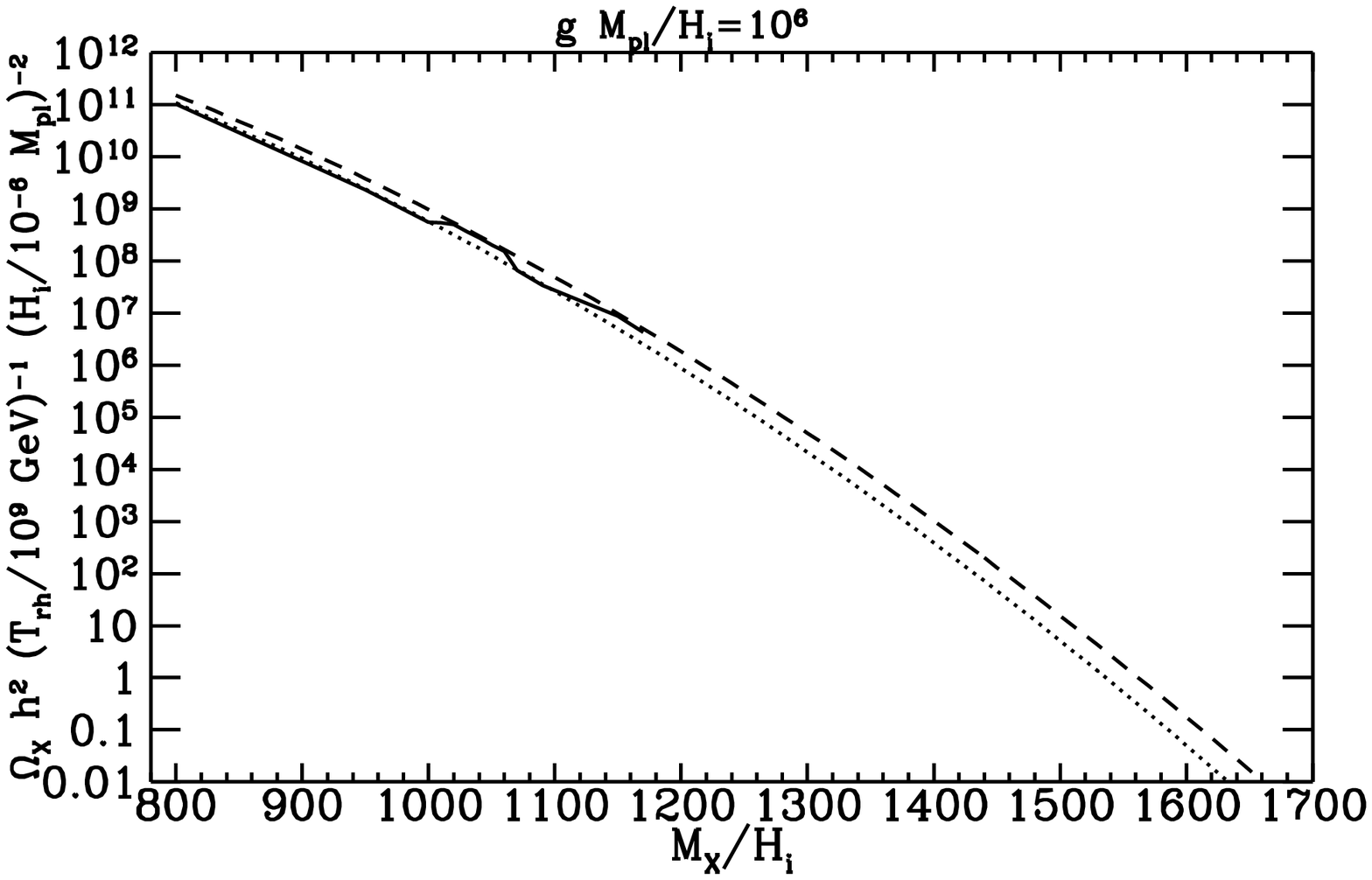}
\caption{The solid curve in this plot of $\omhsq$ versus $b$ for
$\gparam= 10^6$ shows the numerical results.  The correction factor
(explained in the text) used for the dotted curve is 0.932 and for the
dashed curve is 0.903.  Hence, unless if $H_i >10^{-6} \mpl$ or
$T_{\mbox{RH}} > 10^9$ GeV, cosmologically significant dark matter
production in this scenario does not occur for masses above about
$1700 H_i$, which can be in the GUT scale.}
\label{fig:g1e6}
\end{figure}

\begin{figure}[t]
\hspace*{25pt} \epsfxsize=400pt \epsfbox{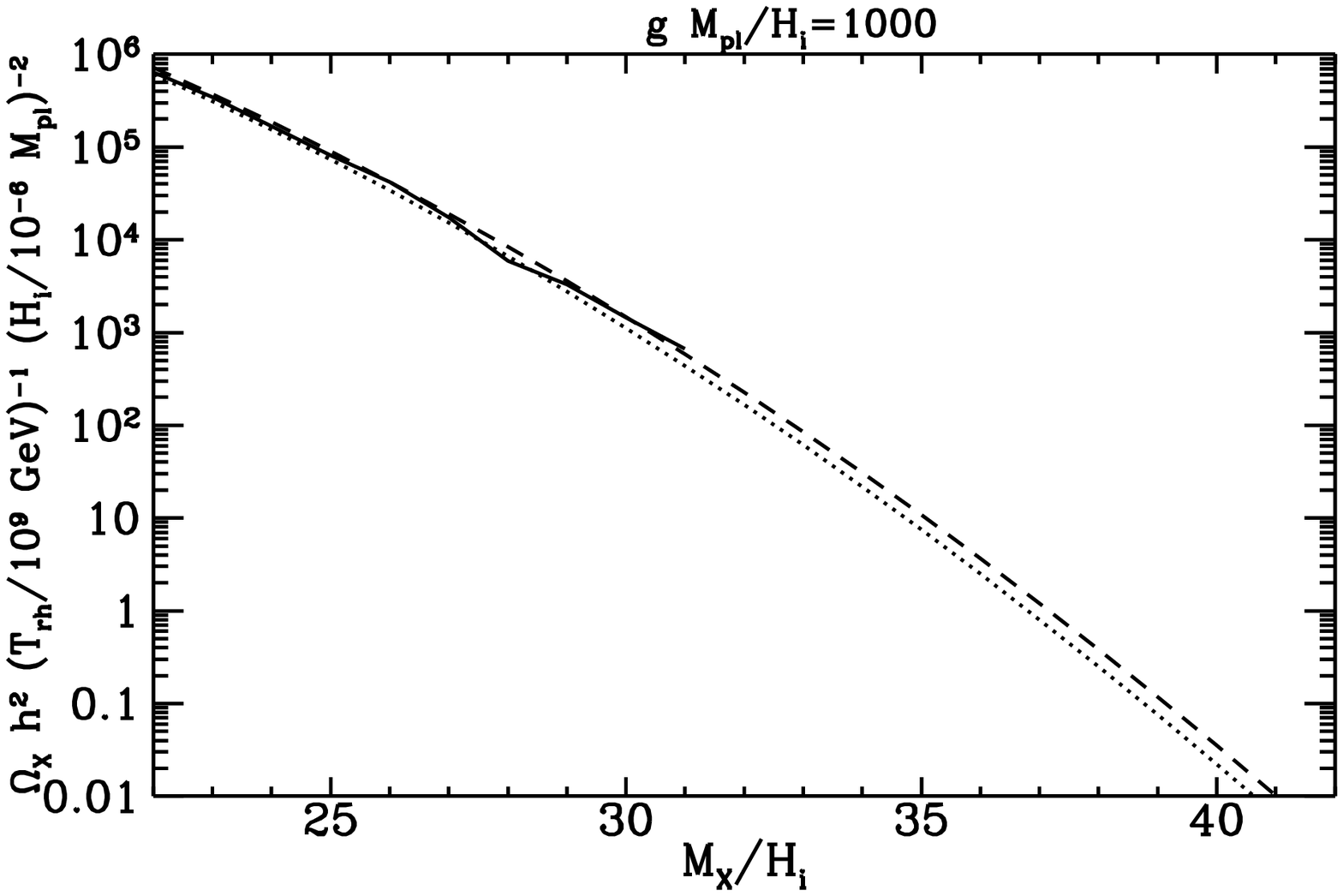}
\caption{The solid curve in this plot of $\omhsq$ versus $b$ for
$\gparam= 10^3$ shows the numerical results.  The correction factor
(explained in the text) used for the dotted curve is 0.98 and for the
dashed curve is 0.963, both of which are quite close to the values used
in Fig.\ \ref{fig:g1e6}.  Hence, we see that our analytic
approximation seems to be in agreement with the numerical results.
}
\label{fig:g1000}
\end{figure}

Because $|\beta_k|^2$ required to have $\omhsq \sim 1$ is extremely
small and one must in general compare oscillatory functions to obtain
it, it is difficult to calculate it numerically.  For $\omhsq \sim 1$,
one requires an accurate calculation of
\be
|\beta_{\bar{k}}|^2 \sim \frac{10^{-5}}{b^3 (\bar{k}/(a_i H_i))^3}
\ee
where $\bar{k}/(a_i H_i)$, the average momentum component, is
typically around $0.3 b$.   Hence, for $b \sim 10^6$, even with
appropriate scaling, the calculation is numerically delicate.
Furthermore, since many momentum components must be calculated for the
integration of the spectrum to obtain the number density, an accurate
calculation is time consuming, at least within a straight forward
framework of calculation.
The results
presented in this section comes from a Runge-Kutta 
solution to a system of equations including Eqs.\ (\ref{eq:modeproper}),
(\ref{eq:friedmann}), and (\ref{eq:phievolve}), all appropriately
scaled both in the independent and the dependent
variables.

\begin{figure}[t]
\hspace*{25pt} \epsfxsize=400pt \epsfbox{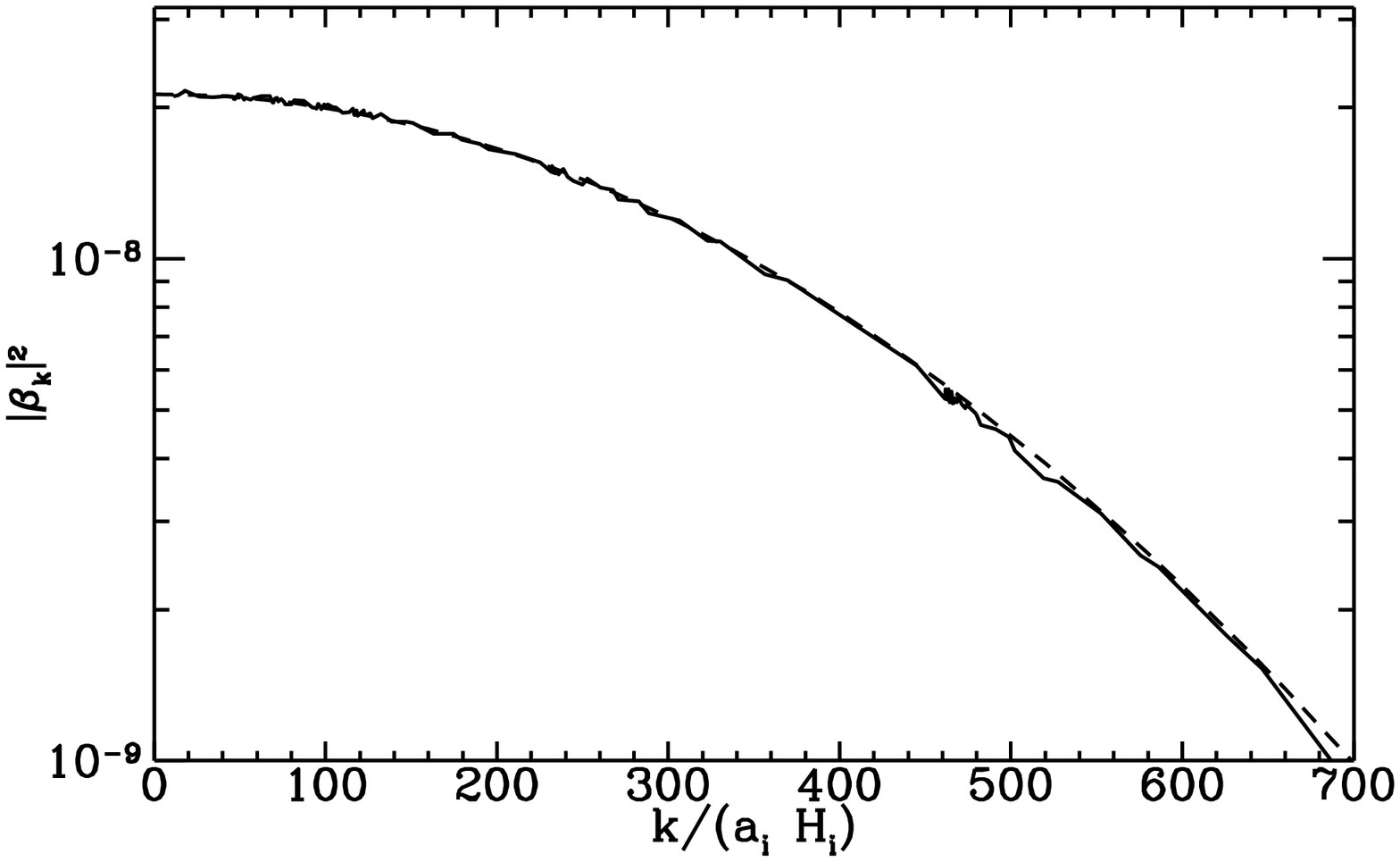}
\caption{For the parameters $\gparam=10^6$ 
and $b=1080$, the particle
density per mode is plotted as a function of the 
wave vector scaled by the scale factor $a_i$ and the Hubble expansion
rate $H_i$ at the end of inflation.  The solid curve corresponds to the
the brute force numerical results.  The dashed curve corresponds to
the spectrum obtained using the analytic approximation
\protect{\eqr{eq:correctedapprox}} with $f=0.932$.
}
\label{fig:speccompare}
\end{figure}

\begin{figure}[t]
\hspace*{25pt} \epsfxsize=400pt \epsfbox{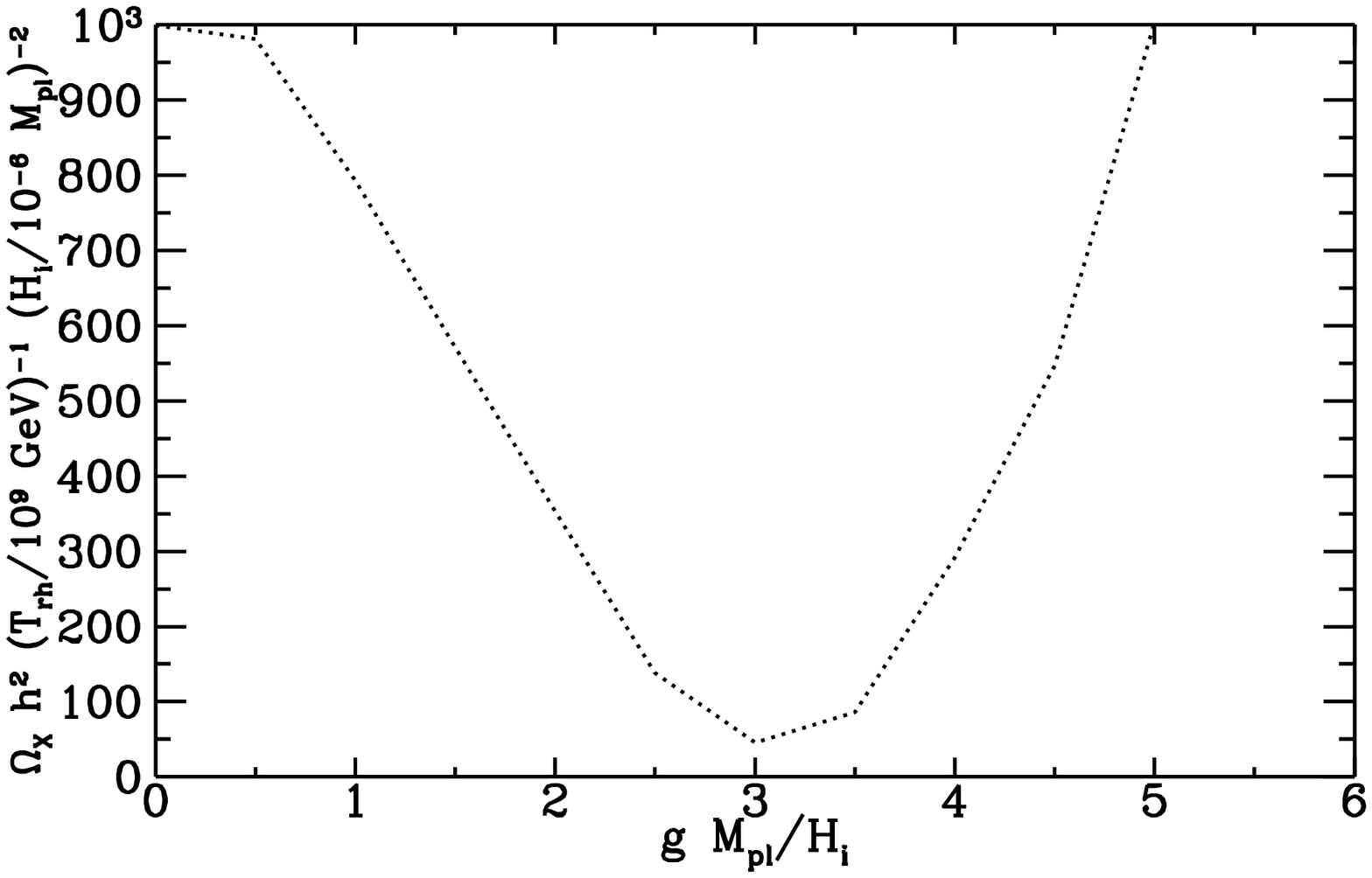}
\caption{ The plot shows the nonmonotonic behavior of the particle
density produced with the variation of the coupling constant.  The
value of $b\equiv M_X/H_i$ is set to 1 here.  
}
\label{fig:cancellation}
\end{figure}

A sample of the numerical results are presented in Figs.\
\ref{fig:g1e6} and \ref{fig:g1000}.  Motivated by our experience with
the comparison with exactly soluble cases in section IV, where we have
seen that the Taylor expansion approximation used gets the exponent
correct only up to a constant multiplicative factor close to 1, we introduce a
correction factor $f$ in the exponent of \eqr{eq:goodapprox} as
\be
|\beta_k|^2 \approx \exp\left( -4 f \left[ \frac{(k/\aeff)^2}{M_X
\sqrt{\heffsq + 
\reff/6}} + \frac{M_X}{\sqrt{\heffsq+\reff/6}} \right] \right)
\label{eq:correctedapprox}
\ee
and integrate this with respect to $k$ to get $\omhsq$ that is plotted
with the solid curve in Figs. \ref{fig:g1e6} and
\ref{fig:g1000}.
It is clear from the fact that $f$ is about 1 in
both the $\gparam=10^6$ and $\gparam=10^3$ cases that our analytic
approximation is a reasonable one.  Furthermore, as exemplified in
Fig.\ \ref{fig:speccompare},  the fit to any
particular spectrum by adjusting only $f$ gives nearly
an exact fit with $f$ between 0.9 and 1.0.  From Fig.\ \ref{fig:g1e6}, $M_X
\sim 1700 H_i \sim 10^3 m_\phi$ is the maximum possible value of
superheavy dark matter that can be produced in cosmologically
significant abundance in this inflationary scenario unless $H_i \gg
10^{-6} \mpl$ or $T_{\mbox{rh}} \gg 10^9$ GeV.

Before concluding, we would like to note a phenomenon that occurs
when the nonadiabaticity of the mode frequency change caused by the
inflaton coupling is comparable to the nonadiabaticity caused by the
gravitational coupling.  If we crudely approximate the exponent of
\eqr{eq:goodapprox} as
\be
-4 \left[ \frac{(k/\aeff)^2}{M_X \heff} + \frac{M_X}{\heff} \right]
\ee
which is qualitatively reasonable in our scenario, then we can find
using crude estimates made in the Appendix (i.e. $H_i \approx m_\phi/3$,
$\phi \approx \mpl/\sqrt{12 \pi}$, and $\dot{\phi} \approx - \mpl
m_\phi/(2 \sqrt{3 \pi})$) that $\omhsq$ vanishes at $\gparam \approx 4
b$.  Of course, in reality the nonadiabaticity is not perfectly
canceled, and we expect only a dip in the particle density as
$\gparam$ is increased from 0.  Indeed, this cancellation of
nonadiabaticity has been observed numerically as illustrated by Fig.\
\ref{fig:cancellation}.  Hence, in general, for small positive $g^2$
couplings to the inflaton field, the
particle production is not a monotonic function of the coupling
constant because of the presence of the classical gravitational field.

\vspace{36pt}
\centerline{\bf VII. SUMMARY}
\vspace{24pt}

In this article, we have considered the production of superheavy
particles $X$ that are coupled to a homogeneous classical inflaton
field $\phi$.  We have found that within the context of a reasonable
$V(\phi)= m_\phi^2 \phi^2/2$ slow-roll inflationary scenario with the
coupling $ g^2 X^2 \phi^2/2$, the parametric resonance phenomenon
tends to overproduce the number of dark matter particles.  Only when
the resonance phenomenon completely shuts off is the number density
small enough to be consistent with the constraint $\Omega_X h^2 <1$.
For the superheavy particles to be cosmologically significant today
($\Omega_X \sim 1$), its mass can be as large as about $1000 H_i \sim
1000 m_\phi$ as indicated by Fig.\ \ref{fig:g1e6}.  Taking into account
the COBE normalization of the curvature perturbation power spectrum,
i.e. $2 P_R^{1/2}/5 \approx 1.91 \times 10^{-5}$ at a scale of $k
\approx 7.5 H_0$ (see for example \cite{review}), and using the approximation
\be
\frac{2}{5} P_R^{1/2} = \frac{8^{3/2} \sqrt{\pi}}{\sqrt{75} \mpl^3} 
\frac{V^{3/2}}{V'}
\ee
where the inflaton potential is evaluated at the time of horizon exit
(about 50 e-folds before the end of inflation),
we can estimate $m_\phi \sim 10^{13}$ GeV.  This means that
the mass of the dark matter particles can be of the order of the GUT
scale.

In the process of making the superheavy dark matter mass range
estimate, we have derived a simple general formula \eqr{eq:goodapprox}
giving an estimate of the particle production due to interactions with
classical fields in the limit that there is only one, dominant
nonadiabatic time period during which particle production ``occurs.''
The time dependent quantities in \eqr{eq:goodapprox} can be
approximately evaluated at the point at which $C''(r)/C^2(r)$ (primes
refer to conformal time derivatives) is a maximum, where for example,
for couplings of the form $Q(\phi) X^2/2 + \xi R X^2/2$, $C$ is given
by \eqr{eq:example}.  This result is applicable to almost any case of
time varying homogeneous classical scalar field interacting with a
quantum field as long as the number of particles produced is small.

Finally, we pointed out a phenomenon in which the number of $X$ particles
produced actually decreases as the coupling to the inflaton field
$\phi$ increases (Fig.\ \ref{fig:cancellation}).  This is a simple
consequence of the fact that the nonadiabatic variation of the
inflaton field is canceled out by the gravitational effect of
nonadiabatic change in the scale factor.

As far as the observability of SDM is concerned, the prospects seem no
better than in the case of electroweak scale WIMPs, unless the WIMPs
are strongly interacting or charged.  Previously, charged
\cite{rujula,limit} or strongly 
interacting dark matter \cite{starkman} has been ruled out with a
combination of 
arguments coming from the unitarity bound \cite{griestkam} and
experimental observations.  Since the SDM evades the usual unitarity
bound, charged or strongly interacting dark matter may still be viable
in this scenario.  

If the SDM decays via an electromagnetic or a hadronic decay, its
decay products may change the spectrum of the diffuse gamma ray
background \cite{kribs}.  Hence, the diffuse background photon
measurements of EGRET and COMPTEL give strong constraints to such
decaying particles for a fairly large range of life times.  Note that
Refs.\ \cite{kuzmin,kr,bkv} also suggest that if the SDM decays, its
decay products may manifest in the form of ultrahigh energy cosmic
rays.  Note that in these decay scenarios, the life time must be in
general much longer than the age of the universe.  For example,
according to Ref.\ \cite{bkv}, if the cosmic ray energy spectrum above
$10^{11}$ GeV is produced by the decay of SDM, the lifetime of the SDM
must be around $10^{12} \xi_X$ in units of the age of the universe
where $\xi_X$ is the fraction of the cold dark matter in SDM.

In general, we do not expect the direct WIMP detection experiments to be
sensitive to SDM because of their low abundance.  The
detection rate which goes like $ R \sim \rho_0 \sigma v/(M_X m_N)$
where $\rho_0$ is the matter density of the halo, $\sigma$ is the
elastic scattering cross section, $v$ is the virial speed, and  $m_N$
is the mass of the nucleon.  Hence, unless if the particles are
strongly interacting in which case $\sigma \sim 10 \mbox{mb}$
\cite{mohapatra}, the WIMP detectors will not have sufficient event
rates to measure these particles.

The indirect method of dark matter detection (detecting the energetic
neutrinos produced by annihilation of dark matter captured through
elastic collisions \cite{jkg}) will also have difficulty in the SDM
scenario.  The neutrino detection rate in general
depends upon the SDM's capture rate in the sun through elastic
collisions, its annihilation rate, and the neutrino cross section for
the production of leptons in the rock or the detector.  Since SDM mass
will be much greater than that of the elastic scatterer, it will lose
very little of its momentum per elastic collision (fractionally
$m_N/M_X$ where $m_N$ is the mass of the nucleon).  Hence, in addition
to the small number density ($0.4 ~\mbox{GeV}/\mbox{cm}^3/M_X$) to begin with,
the capture probability through elastic collisions will be negligible.
The annihilation rate will also be suppressed (unitarity bound $\sim
1/M_X^2$) even if one assumes maximal branching fraction to the
neutrino producing channels.  However, the the cross section for the
production of leptons in the rock or the detector will be
significantly enhanced.  Still, because the cross section will only
grow like $\sqrt{M_X}$ for $M_X$ much greater than the mass of
$W^{\pm}$ (assuming that the neutrinos will have energies that scale
like $M_X$), the enhancement will not be sufficient to overcome the
suppressions.  Indeed, even if one neglects the neutrino absorption
rate in the sun, if there is no significant accretion of SDM in the
sun, one can easily show that the detection
rate will be much smaller than the current detector sensitivity
\cite{jkg} of $10^{-2} \mbox{m}^{-2} \mbox{yr}^{-1}$.

Because the dark matter searches have focused mainly on those
particles with masses that are less than about 100 TeV, the
observational consequences for SDM have been relatively unexplored.
Since as shown in Refs.\ \cite{ckr1,ckr2} and in this paper,
production mechanisms exist for such particles, and since Refs.\
\cite{gaugemed2,hamaguchi,benakli} has shown such particles exist in
extensions to the standard model, a more careful study of
observational consequences of SDM scenarios may be worthwhile.


\vspace{36pt}
\centerline{\bf ACKNOWLEDGMENTS}
\vspace{24pt}

I thank Rocky Kolb for suggesting the problem and commenting
insightfully regarding the manuscript.  I also thank Michael Turner,
Simon Swordy, and Emil Martinec for their comments on this work.  This
research was supported by the DOE and NASA under Grant NAG5-7092.

\frenchspacing
\def\prpts#1#2#3{Phys. Reports {\bf #1}, #2 (#3)}
\def\prl#1#2#3{Phys. Rev. Lett. {\bf #1}, #2 (#3)}
\def\prd#1#2#3{Phys. Rev. D {\bf #1}, #2 (#3)}
\def\plb#1#2#3{Phys. Lett. {\bf #1B}, #2 (#3)}
\def\npb#1#2#3{Nucl. Phys. {\bf B#1}, #2 (#3)}
\def\apj#1#2#3{Astrophys. J. {\bf #1}, #2 (#3)}
\def\apjl#1#2#3{Astrophys. J. Lett. {\bf #1}, #2 (#3)}
\def\pla#1#2#3{Phys. Lett. {\bf #1A}, #2 (#3)}

\vspace{0.25in}



\vspace{36pt}
\centerline{\bf APPENDIX}
\vspace{24pt}
\setcounter{equation}{0}
\renewcommand{\theequation}{A\arabic{equation}}

In this appendix we remind the reader of the boundary
conditions for the classical fields (i.e. \eqr{eq:friedmann}  and
\eqr{eq:phievolve} ) which is 
solved numerically.  As 
we mentioned earlier, we will choose the 
initial conditions as to set up a slow-roll, large-field inflationary
epoch, at the end of which the particle creation will occur.  To gain
intuition for the needed initial conditions, consider the slow-roll
scenario as presented in \cite{KT}.  From the Friedmann equation, 
\be
H \approx \sqrt{\frac{8 \pi}{3 \mpl}}
\frac{\sqrt{V(\phi)}}{\mpl}
\label{eq:potdom}
\ee
where $\dot{\phi}^2 \ll V(\phi)$ has been assumed (slow-roll scenario)
and $V(\phi)$ is the inflaton potential.   Since in general
$\ddot{\phi} \ll 3 H \dot{\phi}$ is required for 
sufficient inflation to occur, we can write
\be
3 H \dot{\phi} \approx -V(\phi)_{,\phi}.
\label{eq:inffried}
\ee
Combining this with \eqr{eq:potdom} gives
\be
a \approx a(t_p) \exp\left( \frac{- 8 \pi}{\mpl^2}
\int_{\phi(t_p)}^{\phi(t)} \frac{V(\phi)}{V_{,\phi}(\phi)} d\phi \right)
\label{eq:efold}
\ee
where $t_p$ ($p$ stands for past) is the time at the beginning of
inflation.  Hence, for potentials of the form $\mbox{constant} \times
\phi^n$, the number of e-folds of inflation is determined solely by the 
initial and the final values of $\phi$.

The end of the slow-roll scenario is obtained by determining when the
potential energy of the inflaton becomes comparable to the kinetic
energy.  \eqr{eq:potdom} and \eqr{eq:inffried} combine to give (in
addition to \eqr{eq:efold})
\be
\dot{\phi}= -\frac{\mpl V_{, \phi} }{\sqrt{24 \pi} \sqrt{V(\phi)}}.
\label{eq:phidotandv}
\ee
which when combined with the approximate condition for the end of
inflation (namely $\dot{\phi}^2/2 \approx V(\phi)$), we have
\be
\frac{V(\phi(\tend))}{V_{,\phi}(\phi(\tend))} \approx
\frac{\mpl}{\sqrt{48 \pi}}
\ee
where $\tend$ is the time at the end of inflation.

Therefore, for potentials of the form $ \mbox{constant} \times \phi^n$,
fixing the number of e-folds fixes the initial value.  For $V(\phi) =
m_\phi^2 \phi^2/2$ potential, we thus have
\be
\phi(\tend) \approx \mpl/\sqrt{12\pi}
\ee
and
\be
\phi(t_p) \approx \frac{\mpl}{\sqrt{2  \pi}} \sqrt{ N_e + 1/6}
\ee
where $N_e$ is the number of inflationary e-folds.  For our typical
runs, we use $N_e \approx 64$.  Then \eqr{eq:phidotandv} (equivalent
to
assuming $\ddot{\phi}=0$) gives the initial condition for
$\dot{\phi}$.

\end{document}